\documentclass[a4paper,11pt]{article}
\usepackage{jinstpub}
\usepackage{hyperref}
\usepackage{orcidlink}
\usepackage{comment}
\usepackage{caption}
\usepackage[toc,page]{appendix}
\usepackage{csquotes}
\usepackage{subcaption}
\usepackage{amsmath}
\usepackage{graphicx}
\usepackage{makecell}
\usepackage[symbol]{footmisc}
\usepackage[font=small,labelfont=bf]{caption}
\title{Optical characterization and ionizing-radiation response of the BCF-20XL scintillating fiber}

\author[a,*]{W.~Bae\orcidlink{0000-0002-7646-7577}\note[{*}]{Corresponding author.},}
\author[a]{L.~Baker\orcidlink{0009-0006-5512-642X},}
\author[a]{K.~Chen\orcidlink{0009-0008-5642-7624},}
\author[a]{J.~Cho\orcidlink{0009-0004-7530-7231},}
\author[a]{K.~Lang\orcidlink{0000-0003-1269-7223},}
\author[a]{C.~Lee\orcidlink{0009-0003-5915-3642},}
\author[a]{E.~Liang\orcidlink{0009-0008-6995-3842},}
\author[a]{C.~Mathurin\orcidlink{0009-0004-0438-3640},}
\author[a]{C.~Murthy\orcidlink{0000-0001-9044-7946},}
\author[a]{D.~Myers\orcidlink{0000-0001-8402-7240},}
\author[a]{S.~Nguyen\orcidlink{0009-0009-9622-6429},}
\author[a]{D.~Phan\orcidlink{0000-0002-0649-8167},}
\author[a]{M.~Proga\orcidlink{0000-0002-0303-5159},}
\author[a]{M.~Zalikha\orcidlink{0009-0002-7045-6022},}
\author[a]{J.~Zey\orcidlink{0009-0008-3005-6642}}
\affiliation[a]{Department of Physics, University of Texas at Austin,\\ 
1 University Station, Austin, TX 78712-0264, USA}
\emailAdd{wonseokb@utexas.edu}

\abstract{We report optical characterization and ionizing-radiation measurements of the BCF-20XL scintillating fiber from Luxium Solutions, which also wavelength-shifts externally incident blue light (referred to here as Sci-WLS).
Attenuation lengths were obtained from transmitted emission spectra using a 3.0\,m fiber with a spectrophotometer and described with a two-component attenuation model, yielding $\lambda_{\text{long}} = 6.70$\,m and $\lambda_{\text{short}} = 0.16$\,m.
The spectral measurements also reveal wavelength-dependent attenuation, with the long component increasing with wavelength and the short component becoming less significant at longer wavelengths.
The scintillation response was also characterized under radioactive $\alpha$, $\beta$, and $\gamma$ irradiation using SiPM-based readout, providing a benchmark of the detected photoelectron response of BCF-20XL.
}


\begin{document}
\maketitle
\flushbottom


\section{Introduction}
\label{sec:introduction}

Plastic scintillating and wavelength-shifting (WLS) fibers have been employed in particle- and nuclear-physics detectors since the 1980s~\cite{Pla-Dalmau:2001, Ruchti:1996}.
WLS fibers have been used in experiments such as MINOS~\cite{MINOS-Michael:2008bc, Avvakumov:2005ww}, NOvA~\cite{Ayres:2004js}, GERDA~\cite{Ackermann:2012xja}, and LEGEND-200~\cite{LEGEND:2025insdet}, while scintillating fibers have been deployed in the D\O{}~\cite{Smirnov:2009} and the LHCb SciFi Tracker~\cite{LHCb-SciFi-NIMA2025}.

The use of a primary scintillating fluor together with a secondary wavelength-shifting fluor in scintillating fibers is well established~\cite{Ruchti:1996}.
In this work, the term scintillating-wavelength-shifting (Sci-WLS) is used descriptively for fibers that produce scintillation under ionizing radiation and also absorb and wavelength-shift incident visible light within their absorption band~\cite{Bae:2025fiberLED, Bae:2025ionRad}.
A potential application is the self-tagging of backgrounds originating in fibers in low-background veto systems, including configurations considered for the proposed LEGEND-1000 detector~\cite{LEGEND:2021bnm,Bae_PhDthesis}.
For example, a decay of \(^{214}\mathrm{Bi}\) or \(^{208}\mathrm{Tl}\) within a fiber produces scintillation through its emitted \(\beta^{-}\) particle, while an accompanying \(\gamma\) ray deposits energy in a nearby germanium detector.
The resulting coincidence could identify the event as originating in the fiber~\cite{Bae_PhDthesis}.

\begin{figure}[t]
  \centering
  \includegraphics[width=0.80\linewidth]{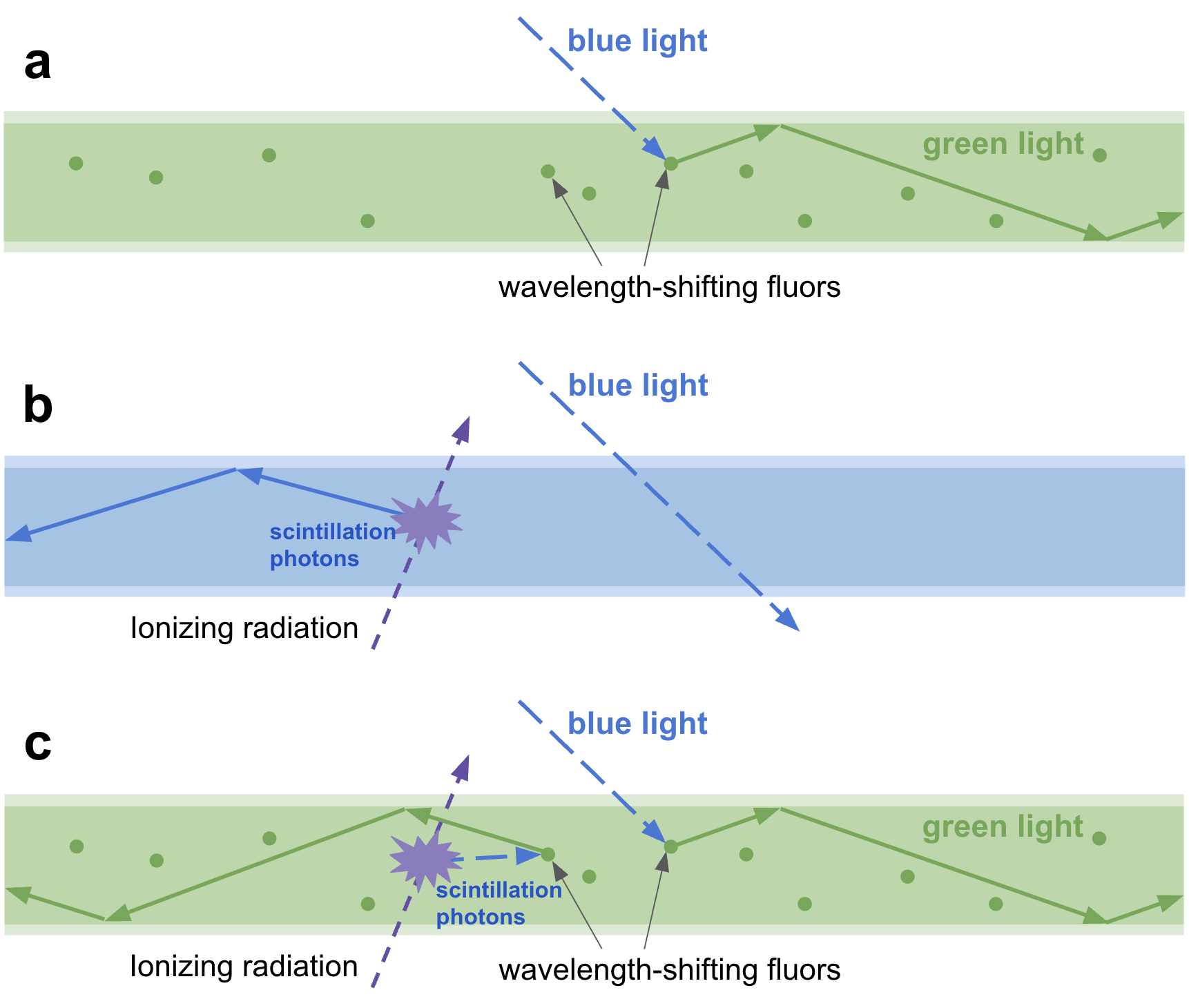}
  \caption{
  Conceptual comparison of three fiber types.
  (a) A WLS fiber: external blue photons are absorbed and re-emitted at longer wavelengths by WLS fluors in the core.
  (b) A blue-emitting scintillating (Sci) fiber: ionizing radiation directly produces scintillation light in the core, while the fiber is largely transparent to externally incident blue light.
  The dashed blue ray illustrates untrapped transmission under side illumination; this is discussed further in appendix~\ref{app:side_incident_blue}.
  (c) A dual-function Sci-WLS fiber: external blue photons can be wavelength shifted as in (a), while scintillation light produced by ionizing radiation is also absorbed and re-emitted by the same WLS fluors.
  In all cases, the produced or re-emitted light is guided to the fiber ends by total internal reflection.
  This figure is adapted from~\cite{Bae_PhDthesis}.}
  \label{fig:WLS_vs_SciWLS_guiding}
\end{figure}

Figure~\ref{fig:WLS_vs_SciWLS_guiding} illustrates selected light-collection mechanisms in WLS, scintillating, and Sci-WLS fibers.
In a WLS fiber, external blue light can be absorbed and re-emitted at longer wavelengths.
In a scintillating fiber, ionizing radiation produces light in the core.
In a fiber with both responses, externally incident blue light can be wavelength-shifted, while light produced by ionizing radiation can also be absorbed and re-emitted by the same WLS fluors.

We focus on the BCF-20XL scintillating fiber manufactured by Luxium Solutions (formerly Saint-Gobain)~\cite{Luxium-Solutions, Saint-Gobain2}.
We characterized the LED-based optical response and ionizing-radiation-induced scintillation response of the BCF-20XL fiber, establishing performance benchmarks using the experimental and analysis procedures developed in our previous fiber-characterization studies~\cite{Bae:2025fiberLED, Bae:2025ionRad}.

\section{Fiber samples}
\label{sec:fiber_samples}

\begin{figure}[h!]
    \setlength{\abovecaptionskip}{5pt}   
    \setlength{\belowcaptionskip}{0pt}   
    \centering
    \includegraphics[width=.495\textwidth, height=.37\textwidth]{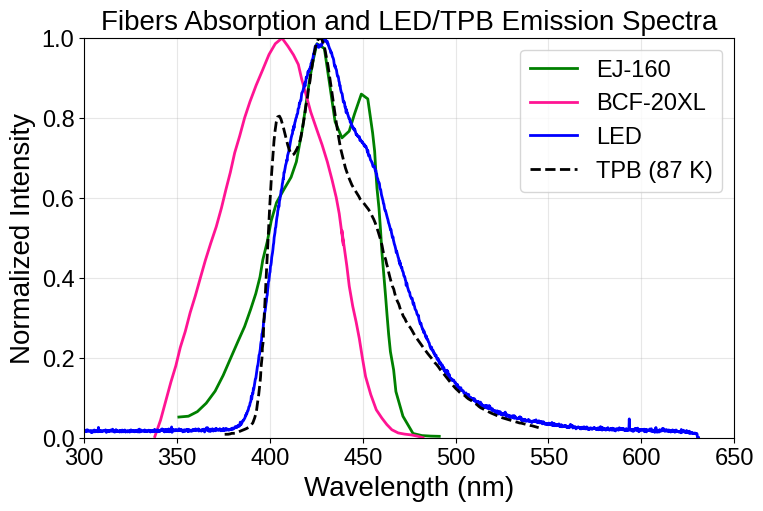}
    \includegraphics[width=.495\textwidth, height=.37\textwidth]{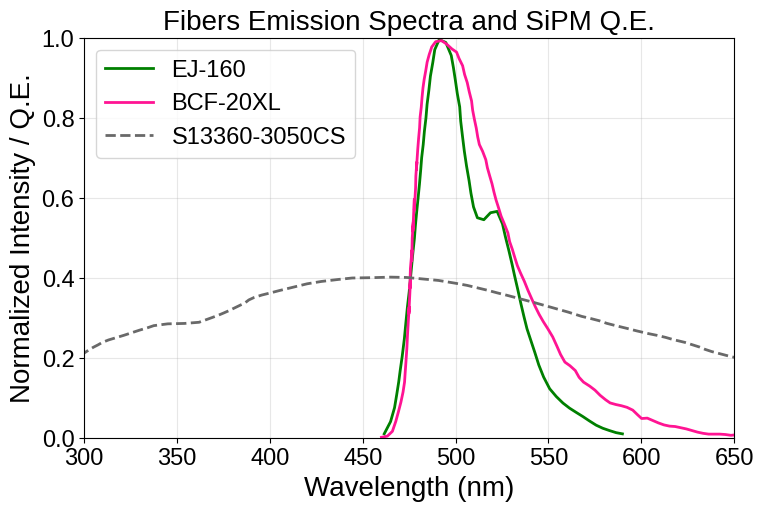}
    \caption{Absorption (left) and emission (right) spectra of the BCF-20XL fiber 
    (manufacturer data~\cite{Luxium-Solutions}). 
    The left panel also shows the blue LED spectrum used for optical excitation, selected to overlap the TPB emission spectrum at 87 K shown alongside it~\cite{TPB-Leonhardt-JINST-2024}.
    For comparison, the Sci-WLS fiber EJ-160 spectra from Eljen Technology~\cite{Eljen2} and the quantum efficiency of the Hamamatsu S13360 SiPM~\cite{hamamatsu} are included. 
    All spectra are normalized to their respective maxima. 
    The LED, TPB, EJ-160, and SiPM reference curves are adapted from~\cite{Bae:2025fiberLED}.}
\label{fig: fiber emission spectrum1}
\end{figure}

\begin{figure}[h!]
\centering
\includegraphics[width=.35\textwidth]
{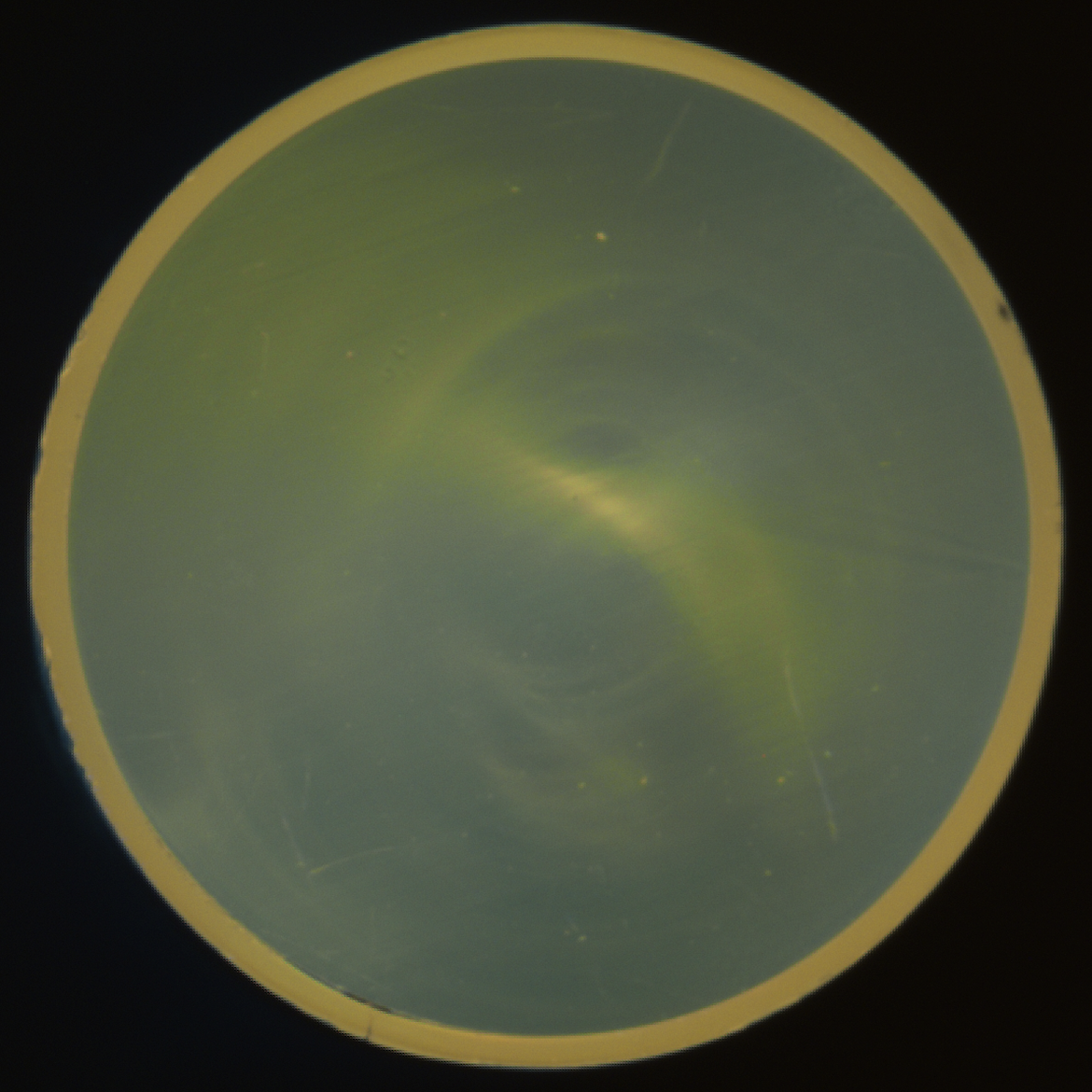}
\caption{
Microscope cross-sectional view of the diamond-fly-cut tested fiber. External illumination was applied to clearly reveal the core-cladding boundary.
The bright pattern visible within the fiber core is a reflection artifact from the external illumination, not a structural feature.
}
\label{fig: microscope inspection}
\end{figure}

The main geometric and material properties of the BCF-20XL fiber characterized in this work are summarized in table~\ref{table:wls_fibers1}.
Manufacturer-provided absorption and emission spectra are shown in figure~\ref{fig: fiber emission spectrum1}.
For context, Luxium reports a light yield of approximately \(8000\) photons/MeV for minimum-ionizing particles and an attenuation length greater than \(4\,\mathrm{m}\) for a \(1\,\mathrm{mm}\)-diameter fiber~\cite{Luxium-Solutions}.
Figure~\ref{fig: microscope inspection} shows a representative microscope image of the diamond-fly-cut cross section of the tested fiber, obtained with a Nikon SMZ1500 microscope~\cite{Nikon-microscpoe}.

\begin{table}[h!]
\caption{Properties of the tested BCF-20XL fiber. The cladding thickness was measured using a microscope.}
\small
\centering
\begin{tabular}{|p{3.5cm}|c|}
    \hline
    \multicolumn{1}{|c|}{\textbf{Feature}} 
        & \textbf{BCF-20XL} \\
    \hline
    Type 
        & scintillating fiber\\
    \hline
    Cross-section 
        & 1~mm round \\
    \hline
    Cladding   
        & single \\
    \hline
    Core material  
        & polystyrene \\
    \hline
    Cladding material 
        & polymethyl methacrylate \\
    \hline
    \makecell[{{p{3.5cm}}}]{Refractive index \\ (core / cladding)}
        & 1.60/1.49 \\
    \hline
    Cladding thickness 
        & 0.03\,mm\\
    \hline
\end{tabular}
\label{table:wls_fibers1}
\end{table}



\section{Optical characterization}
\label{sec:optical_characterization}


\subsection{Measurement setup}
\label{sec:optical_setup}

Light propagating in an optical fiber can include both core-guided and cladding-guided contributions.
The core-guided component is carried by total internal reflection at the core--cladding boundary and remains relatively stable over long distances, whereas cladding-guided light propagates near the cladding boundary and is lost rapidly, especially within the first meter, because of surface imperfections and boundary conditions.
The cladding-guided contribution can initially exceed the core-guided contribution substantially owing to its larger geometric trapping efficiency~\cite{Amos1990:fiber-cladding_light, Achenbach:fiber-cladding_light}.
Specifically, for isotropic emission within an ideal, air-surrounded circular fiber, the calculated initial trapping efficiency of cladding-guided light is approximately 3--4 times that of core-guided light for typical refractive indices~\cite{Achenbach:fiber-cladding_light}.
Therefore, the distance dependence of the light transport is modeled with two exponential components, where $\lambda_{\text{long}}$ is often associated with the slowly attenuating core-guided contribution and $\lambda_{\text{short}}$ with the rapidly attenuating cladding-guided contribution.

The LED-based optical characterization of the BCF-20XL fiber used the same apparatus and procedure as in our previous studies~\cite{Bae:2025fiberLED}; we summarize the main features here.
The fibers were approximately 3.0\,m long, laid in a spiral groove machined into an acrylic plate (groove diameter 45--60\,cm), ensuring negligible transmission loss due to bending~\cite{Fiber-bending}.
One end was optically coupled to a spectrophotometer~\cite{Ocean-Optics} covering approximately 350--800\,nm with a wavelength precision of about 0.21\,nm and a limited numerical aperture (NA\,$\approx$\,0.22), corresponding to acceptance angles primarily matching the core-guided component; the opposite end was diamond-fly-cut and left uncoated.
Each spectrum was averaged over 10 acquisitions of 20\,ms each. 
A blue LED~\cite{LEDtronics}, shown spectrally in figure~\ref{fig: fiber emission spectrum1}, illuminated the fiber from the side at 20 discrete positions spanning 0.124--2.944\,m from the spectrophotometer input.
All measurements were performed inside a dark enclosure at room temperature ($\sim$21$^\circ$C), with a residual LED-off background subtracted from each measurement.

\subsection{Emission spectra and attenuation lengths}
\label{sec:emission_and_attenuation}

\begin{figure}[h!]
  \centering
  \begin{subfigure}[t]{0.495\textwidth}
    \centering
    \includegraphics[width=\textwidth]{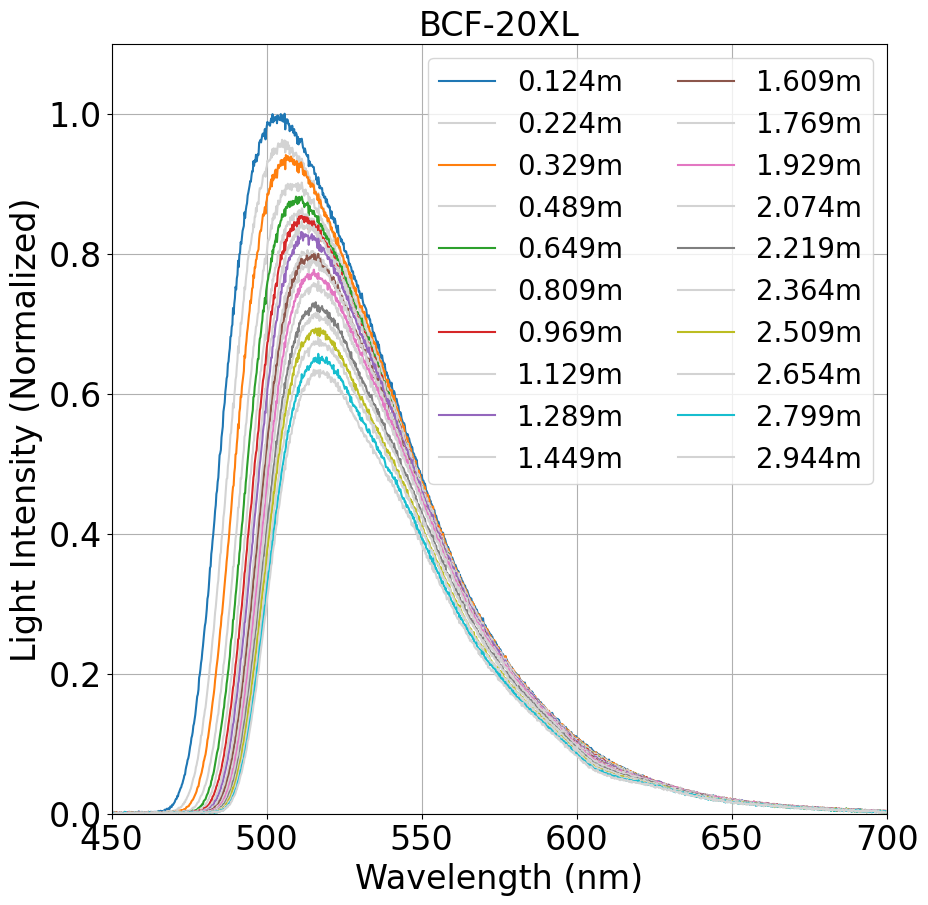}
  \end{subfigure}
  \hfill
  \begin{subfigure}[t]{0.495\textwidth}
    \centering
    \includegraphics[width=\textwidth]{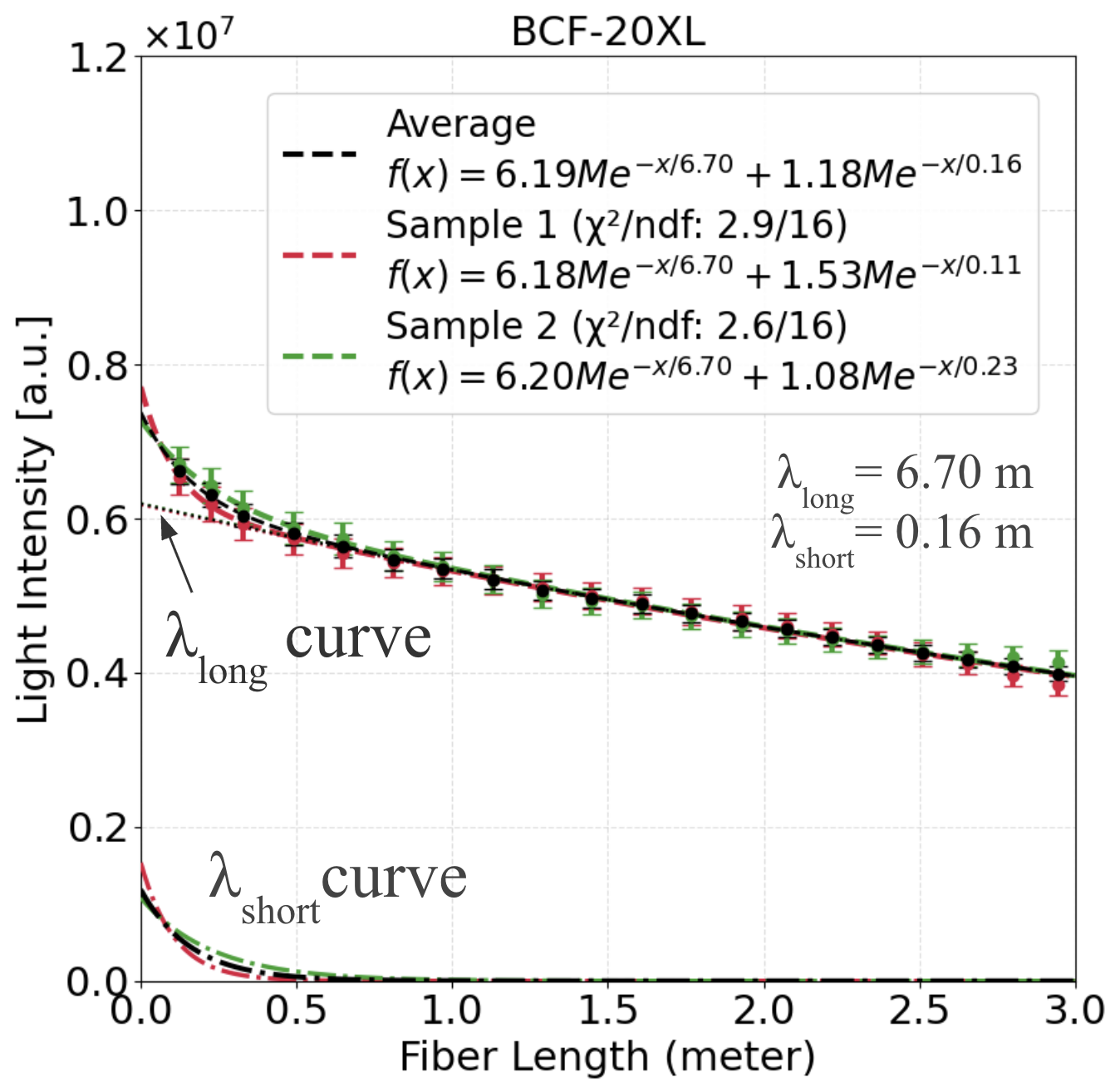}
  \end{subfigure}
\caption{
Optical characterization of the BCF-20XL fiber.
Left: spectra at 20 propagation distances, ordered from shortest to longest (top to bottom near the peak).
Right: spectrally integrated signal as a function of light-propagation distance (fiber length) for two BCF-20XL samples, fitted with the two-component attenuation model (Eq.~\ref{eq:attenuation_doubleexp_optical}).
The long and short components are also shown as dashed lines.}
  \label{fig:bcf20xl_emission_and_attenuation}
\end{figure}

The emission spectra in the left panel of figure~\ref{fig:bcf20xl_emission_and_attenuation} were acquired at 20 LED positions, corresponding to light-propagation distances of 0.124--2.944\,m between the illumination point and the spectrophotometer input.
For visual comparison, all spectra are normalized to the maximum intensity observed at the shortest distance (0.124\,m). 

The right panel of figure~\ref{fig:bcf20xl_emission_and_attenuation} shows the integrated transmitted light intensity for two BCF-20XL samples as a function of propagation distance.
For each spectrum, the light intensity was obtained by integrating the measured emission over 450--700\,nm.
Each point is the mean of 10 consecutive spectrophotometer acquisitions (20\,ms integration each), giving negligible statistical uncertainty. 
A 3.5\% uncertainty per point was assigned from the variation of the integrated emission observed under fiber and LED repositioning used to evaluate measurement repeatability; the fiber was repositioned three times and the LED twenty times, and this value was adopted as a conservative estimate of the measurement repeatability.
For each measurement round, the LED output was tracked with an additional spectrophotometer~\cite{Ocean-Optics}, and the resulting fiber light output was linearly normalized to correct for LED-output variations.

The attenuation of light intensity was modeled as a function of light-propagation distance using a two-component exponential form,
\begin{equation}
I = I_{\text{long}}\,e^{-x/\lambda_{\text{long}}} + I_{\text{short}}\,e^{-x/\lambda_{\text{short}}},
\label{eq:attenuation_doubleexp_optical}
\end{equation}
where $I$ is the spectrally integrated light intensity and $x$ is the 
light-propagation distance; $I_{\text{long}}$, $I_{\text{short}}$, 
$\lambda_{\text{long}}$, and $\lambda_{\text{short}}$ denote the long 
and short components of the light intensities and attenuation lengths.
The integrated-intensity data were fitted with Eq.~\ref{eq:attenuation_doubleexp_optical}.
From the fit, we obtained a long attenuation length of $\lambda_{\mathrm{long}}=(6.70\pm0.08_{\mathrm{fit}})\,\mathrm{m}$ and a short attenuation length of $\lambda_{\mathrm{short}}=(0.16\pm0.02_{\mathrm{fit}} \pm0.07_{\mathrm{sample}})\,\mathrm{m}$, where the sample term is the maximum deviation of the individual-sample fit values from the fit to the intensity averaged across the two samples at each distance (black curve labeled \enquote{Average} in figure~4); $\lambda_{\rm long}$ shows no such deviation.

\subsection{Spectral attenuation lengths}
\label{sec:spectral_attenuation}

The left panel of figure~\ref{fig:bcf20xl_emission_and_attenuation} shows that the spectrum shifts toward longer wavelengths with propagation distance, with short-wavelength components diminishing rapidly while long-wavelength components decrease more slowly.
This evolution of the spectra indicates wavelength-dependent attenuation.
Similar spectral behavior has been observed in other WLS fibers~\cite{Pahlka2019, Bae:2025fiberLED}. 
We quantified this effect by extracting attenuation lengths as a function of wavelength using Eq.~\ref{eq:attenuation_spectral}.

For each narrow wavelength slice, the measured intensity versus light-propagation distance was fitted to a two-component attenuation model,
\begin{equation}
I(\lambda) = I_{\text{long}}(\lambda)\,e^{-x/\Lambda_{\text{long}}(\lambda)} + I_{\text{short}}(\lambda)\,e^{-x/\Lambda_{\text{short}}(\lambda)},
\label{eq:attenuation_spectral}
\end{equation}
where $I(\lambda)$ is the signal at wavelength $\lambda$ and distance $x$, and $\Lambda_{\text{long}}(\lambda)$ and $\Lambda_{\text{short}}(\lambda)$ are the long and short spectral attenuation lengths.

\begin{figure}[h!]
\centering
\includegraphics[width=.495\textwidth, height=.495\textwidth]
{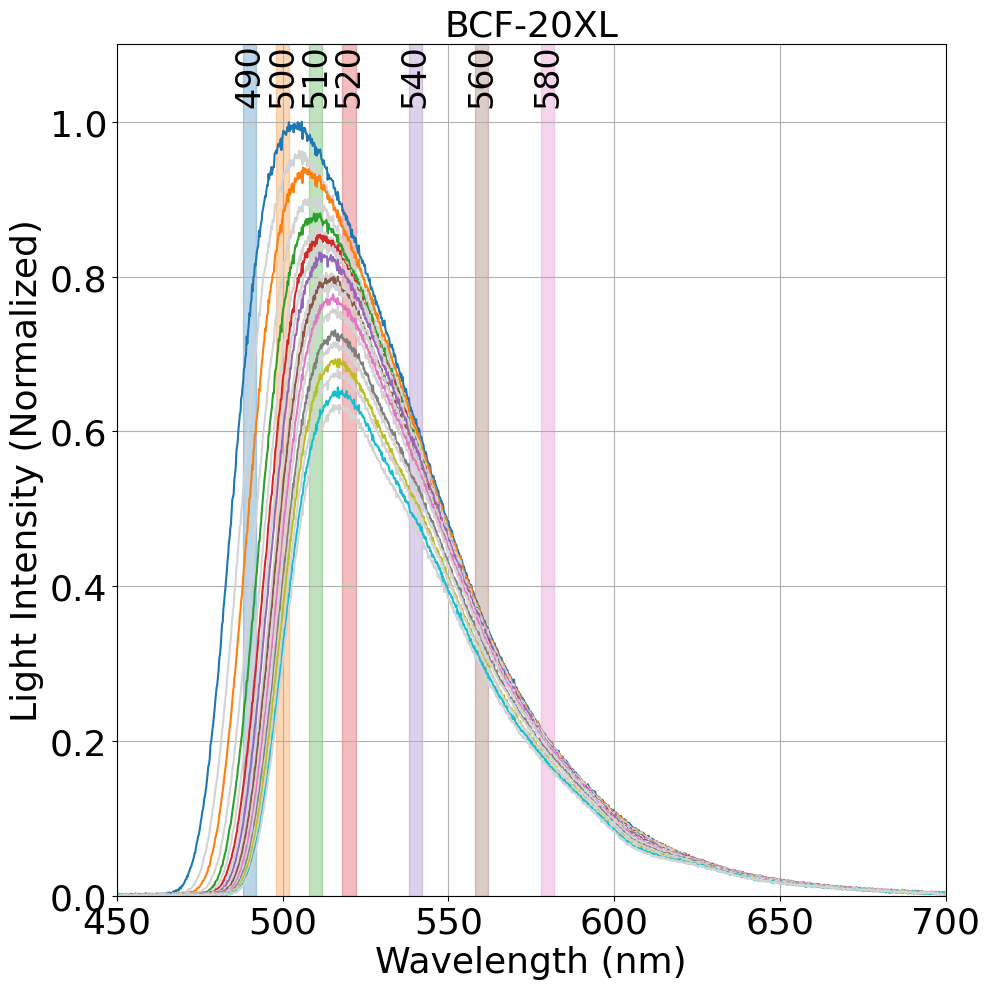}
\includegraphics[width=.495\textwidth, height=.495\textwidth]
{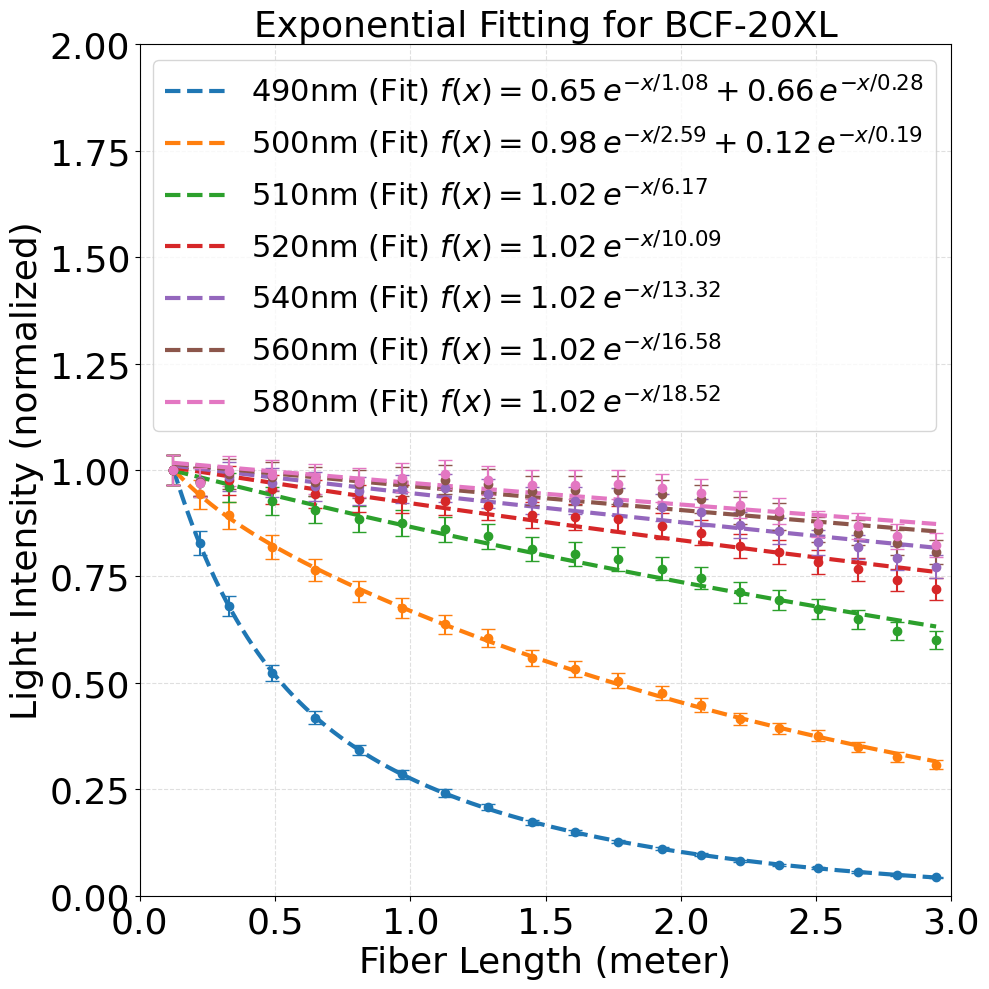}
\caption{\raggedright
(Left) Emission spectra with seven selected wavelength regions indicated. The figure is adapted from figure~\ref{fig:bcf20xl_emission_and_attenuation}.
(Right) Double-exponential fits to each region.}
\label{fig: attenuation length with selected wavelength} 
\end{figure}

We first applied Eq.~\ref{eq:attenuation_spectral} to seven representative wavelength bands shown in figure~\ref{fig: attenuation length with selected wavelength}, using a band half-width of $\pm2.0\,\mathrm{nm}$ to maintain adequate statistics.
To test whether the second component is warranted, we compared single- and double-exponential fits with a $\chi^{2}$ difference test ($\Delta\mathrm{dof}=2$). 
The two-component description is clearly favored around 490--500\,nm, whereas above 510\,nm the improvement is negligible; accordingly, we reported only the long component for $\lambda \gtrsim 510\,\mathrm{nm}$.

\begin{figure}[h!]
\centering
\includegraphics[width=.495\textwidth]{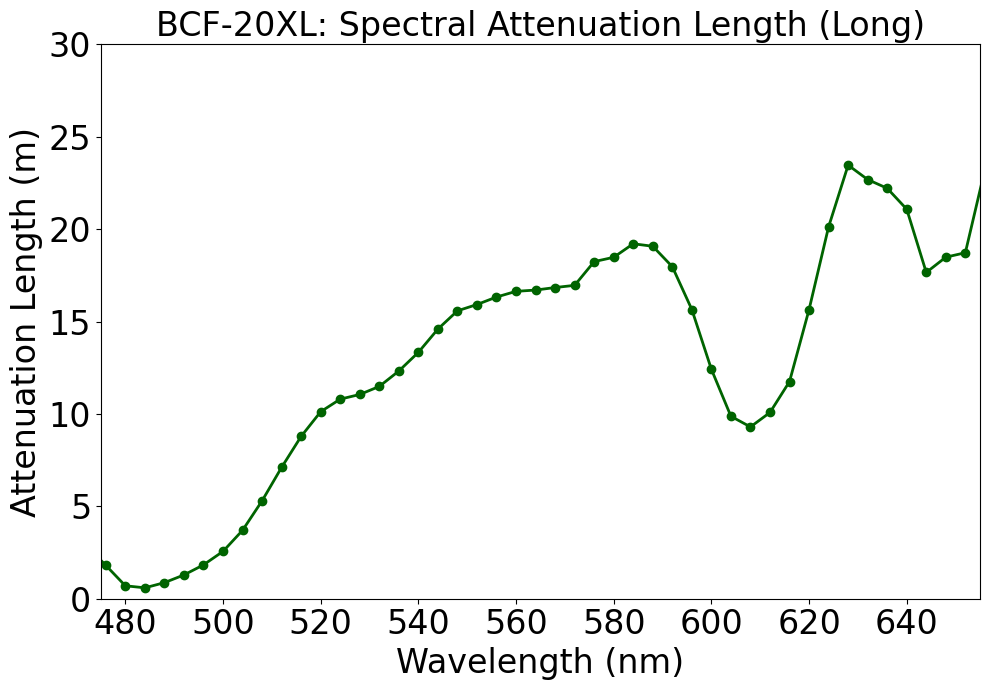}
\includegraphics[width=.495\textwidth]{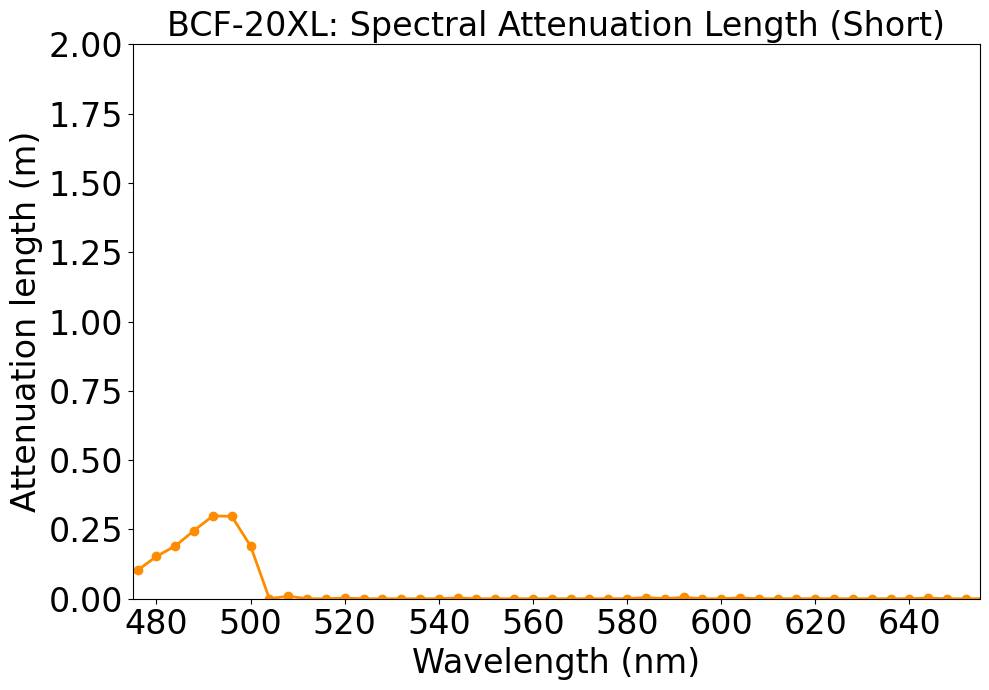}
\caption{Spectral attenuation lengths for BCF-20XL. 
The left column shows $\Lambda_{\text{long}}(\lambda)$ and the right column shows $\Lambda_{\text{short}}(\lambda)$.}

\label{fig: spectral attenuation length}
\end{figure}

The same procedure was then performed over 475--655\,nm, yielding the spectral attenuation curves in figure~\ref{fig: spectral attenuation length}. 
Over this range, $\Lambda_{\text{long}}(\lambda)$ generally becomes larger at longer wavelengths, with local minima near 490, 610, and 650\,nm; this structure is qualitatively consistent with previous WLS-fiber measurements from multiple vendors~\cite{Mu2e-JINST, Bae:2025fiberLED}.
Fits for $\Lambda_{\text{short}}(\lambda)$ can become ill-conditioned where the short contribution is weak; we therefore quote $\Lambda_{\text{short}}(\lambda)$ only when
$ I_{\text{short}}/(I_{\text{long}} + I_{\text{short}}) > 0.03 $,
which stabilizes the extraction and reflects that the short component effectively disappears above the main emission peak ($\gtrsim$505\,nm).


\subsection{Summary of optical characterization}
\label{sec:optical_summary}

In summary, the light attenuation of BCF-20XL was well described by the two-component attenuation model introduced above, and we further extracted the wavelength dependence via $\Lambda_{\text{long}}(\lambda)$ and $\Lambda_{\text{short}}(\lambda)$. 
The long component increases toward longer wavelengths, while the short component becomes negligible above the main emission peak. 
These results provide benchmark attenuation length scales and their spectral dependence for BCF-20XL fiber.


\section{Scintillation response to ionizing radiation}


\subsection{Measurement setup}
\label{sec:irr_setup}


The ionizing-radiation measurements used the irradiation configurations and SiPM-based readout procedure established in our previous study~\cite{Bae:2025ionRad}; we summarize the main features here.
For the $\beta$ and $\gamma$ measurements, an approximately 1.4\,m fiber segment, chosen to match the nominal fiber length in LEGEND-1000~\cite{LEGEND:2021bnm}, was read out at both ends by Hamamatsu S13360-3050CS SiPMs (3$\times$3~mm$^2$ active area, 50\,$\mu$m microcells)~\cite{hamamatsu}.
For the $\alpha$ measurement, fiber segments of different lengths were measured in an end-irradiation configuration with single-end SiPM readout.
All fibers were diamond-fly-cut at both ends, and the SiPM-coupled ends were optically coupled using BC-630 grease~\cite{Saint-Gobain2}.
The SiPMs were mounted on custom readout boards; waveforms were digitized with a Teledyne LeCroy WaveRunner HRO 66Zi oscilloscope~\cite{TeledyneLeCroy}.

In contrast to the spectrophotometer, which accepts light within a limited numerical aperture, the SiPMs collect light from all angles at the fiber end face, thus receiving contributions from both core-guided and cladding-guided light.
As a result, the cladding-guided component dominates at short distances from the photosensor before attenuating rapidly within the first meter~\cite{Amos1990:fiber-cladding_light, Achenbach:fiber-cladding_light}.
The fiber was exposed to standard $\alpha$, $\beta$, and $\gamma$ calibration sources, and the resulting photoelectron (p.e.) count detected by the SiPMs was evaluated as a function of the source-to-SiPM distance.
The measured SiPM pulse amplitudes were converted into detected p.e. counts following the procedure established in our previous study~\cite{Bae:2025ionRad}.
The reported mean detected p.e. counts characterize the scintillation response of BCF-20XL under the specified irradiation and SiPM-readout conditions.
Measurements were performed in a dark enclosure at room temperature ($\sim$21$^\circ$C), with residual source-off backgrounds subtracted from each measurement.


\subsection{Beta response}
\label{sec:Beta_irradiation}

The $\beta$-irradiation response of BCF-20XL was measured using a \textsuperscript{90}Sr point source with an activity of 4.7\,$\mu$Ci.
The $\beta$ measurement used the optical-bench configuration described in~\cite{Bae:2025ionRad}, with a source-collimation arrangement that enabled perpendicular irradiation of the 138\,cm-long fiber at 13 positions spanning 5--133\,cm from one end.



\begin{figure}[h!]
\centering
\captionsetup[figure]{skip=5pt}

\begin{subfigure}{0.49\textwidth}  \includegraphics[width=\linewidth,height=0.65\textwidth]{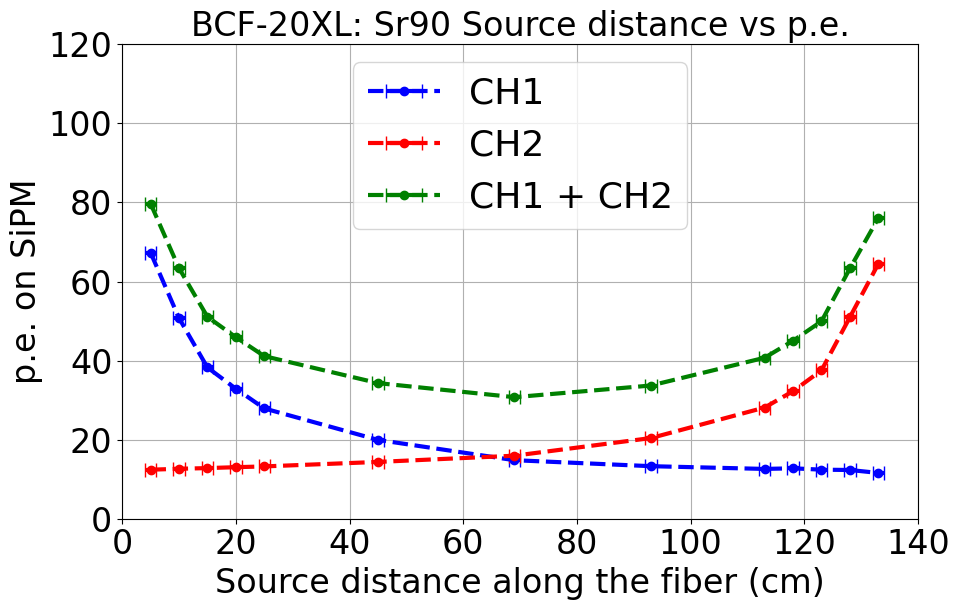}
\end{subfigure}
\hfill
\begin{subfigure}{0.49\textwidth}  \includegraphics[width=\linewidth,height=0.65\textwidth]{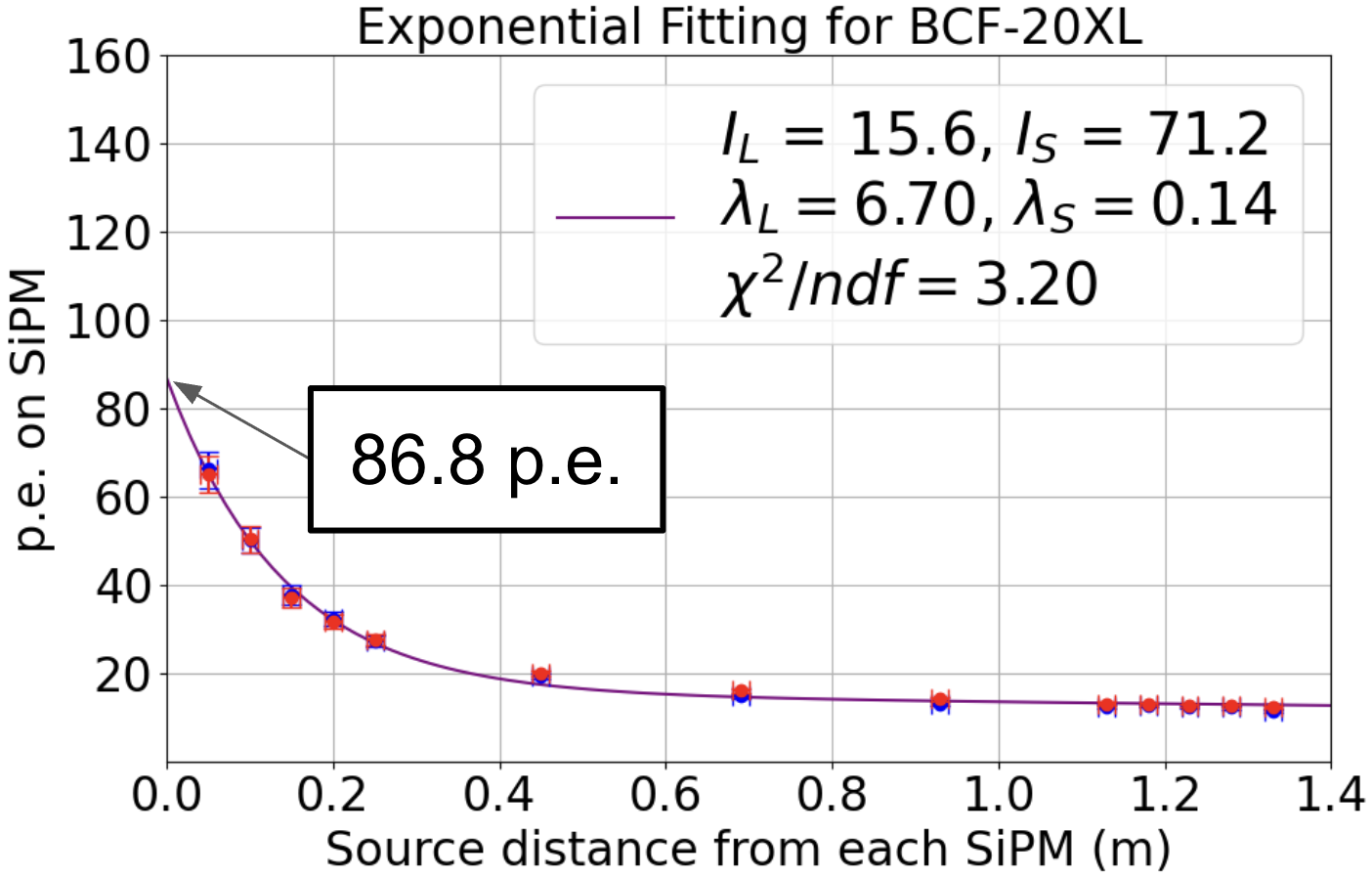}
\end{subfigure}

\caption{Distance dependence of the mean detected p.e. response under \textsuperscript{90}Sr $\beta$ irradiation.
Left: mean detected p.e. count versus source position.
Right: corresponding double-exponential fits.}
\label{fig: beta_sum_fitting}
\end{figure}

At each source position, data were acquired for 10 minutes, yielding tens of thousands of events.
The statistical uncertainty of each mean detected p.e. count, evaluated as the standard error of the mean, is 0.02--0.11, 0.02--0.09, and 0.02--0.10\,p.e.\ for the $\beta$, $\gamma$, and $\alpha$ measurements, respectively, i.e.\ less than 0.25\% of the mean; it is included in the error bars of figures~\ref{fig: beta_sum_fitting}--\ref{fig: alpha_sum_fitting}.

The left panel of figure~\ref{fig: beta_sum_fitting} shows the mean detected p.e. count measured by the left SiPM (CH1), the right SiPM (CH2), and their sum.
The measured values as a function of source position show a symmetric dependence.
The right panel shows the corresponding fits using the same two-component attenuation form as in Eq.~\ref{eq:attenuation_doubleexp_optical}, where $I$ represents the mean detected p.e. count recorded by a given SiPM channel as a function of source position.
The approximately 1.4\,m span of the irradiation measurements was insufficient to constrain $\lambda_{\mathrm{long}}$ independently; it was therefore fixed at the optical value of 6.70\,m obtained in Section~\ref{sec:emission_and_attenuation}.
The fits characterize the distance dependence of the detected response.
Following established practice in scintillation response 
measurements~\cite{Moszynski2013, Moszynski2014, Bae:2025ionRad}, 
we used the fitted functions to extrapolate to $x=0$ for each channel 
to estimate the mean detected p.e. count at $x=0$ under $\beta$ irradiation.
The same fitting prescription was applied to the $\gamma$ and $\alpha$ measurements below.

Systematic uncertainties were evaluated from repeated independent measurements in which the fiber--SiPM coupling and source alignment were reset between trials, including removing and reapplying optical grease.

\subsection{Gamma response}
\label{sec:gamma_irradiation}

The $\gamma$-irradiation response of BCF-20XL was measured using a \textsuperscript{22}Na point source with an activity of 3.0\,$\mu$Ci and a predominant 511\,keV gamma line. 
As in the $\beta$ measurement, the $\gamma$ measurement was performed on the optical bench described in~\cite{Bae:2025ionRad}, with the collimation and shielding arrangement adapted for the \textsuperscript{22}Na source.

\begin{figure}[h!]
\centering
\captionsetup[figure]{skip=5pt}

\begin{subfigure}{0.49\textwidth}
  \includegraphics[width=\linewidth,height=0.65\textwidth]{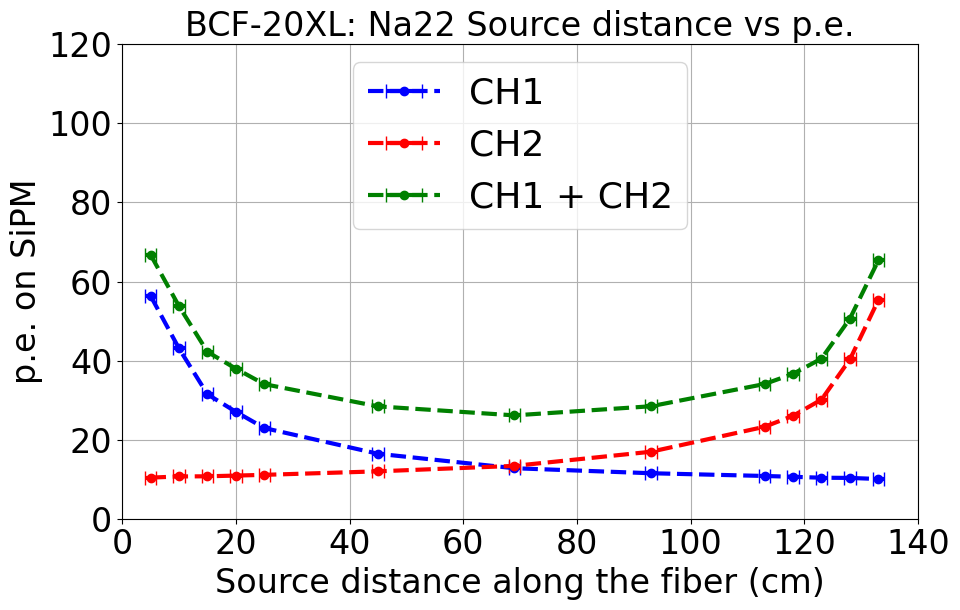}
\end{subfigure}
\hfill
\begin{subfigure}{0.49\textwidth}
  \includegraphics[width=\linewidth,height=0.65\textwidth]{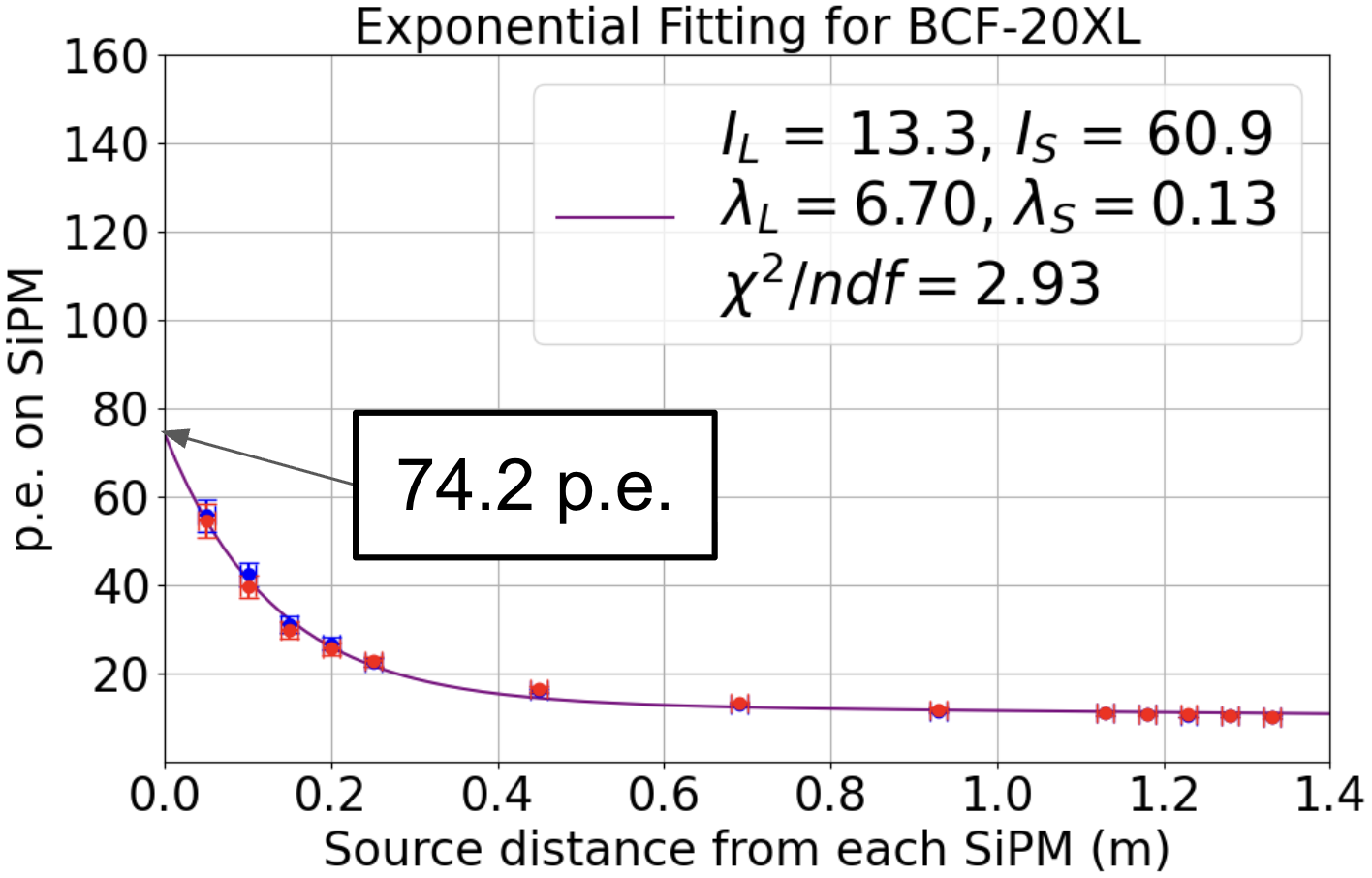}
\end{subfigure}

\caption{Distance dependence of the mean detected p.e. response under \textsuperscript{22}Na $\gamma$ irradiation.
Left: mean detected p.e. count versus source position.
Right: corresponding double-exponential fits.}
\label{fig: gamma_sum_fitting}
\end{figure}

The left panel of figure~\ref{fig: gamma_sum_fitting} shows the mean detected p.e. count measured by the left SiPM (CH1), the right SiPM (CH2), and their sum as a function of the irradiation position. 
The right panel shows the corresponding fits using the same two-component attenuation model introduced in Eq.~\ref{eq:attenuation_doubleexp_optical}. 
The $\gamma$ results follow similar behavior observed in the $\beta$ measurement.

\subsection{Alpha response}
\label{sec:alpha_irradiation}

The $\alpha$-irradiation response of BCF-20XL was measured using a \textsuperscript{241}Am point source with an activity of 1.0\,$\mu$Ci and a predominant $\alpha$ energy of 5.486\,MeV. 
Because $\alpha$ particles have a short range of about 40\,$\mu$m in polystyrene~\cite{Seltzer1993-bq-ASTAR}, 
we adopted an end-irradiation configuration instead of transverse irradiation.
For this measurement, BCF-20XL fiber samples with lengths of 7, 12, 37, 72, and 138\,cm were used in the end-irradiation configuration described in~\cite{Bae:2025ionRad}.
In each measurement, the \textsuperscript{241}Am source irradiated the free end of the fiber, and the other end was optically coupled to one SiPM.

\begin{figure}[h!]
\centering
\captionsetup[figure]{skip=5pt}
  \includegraphics[width=0.5\linewidth,height=0.325\textwidth]{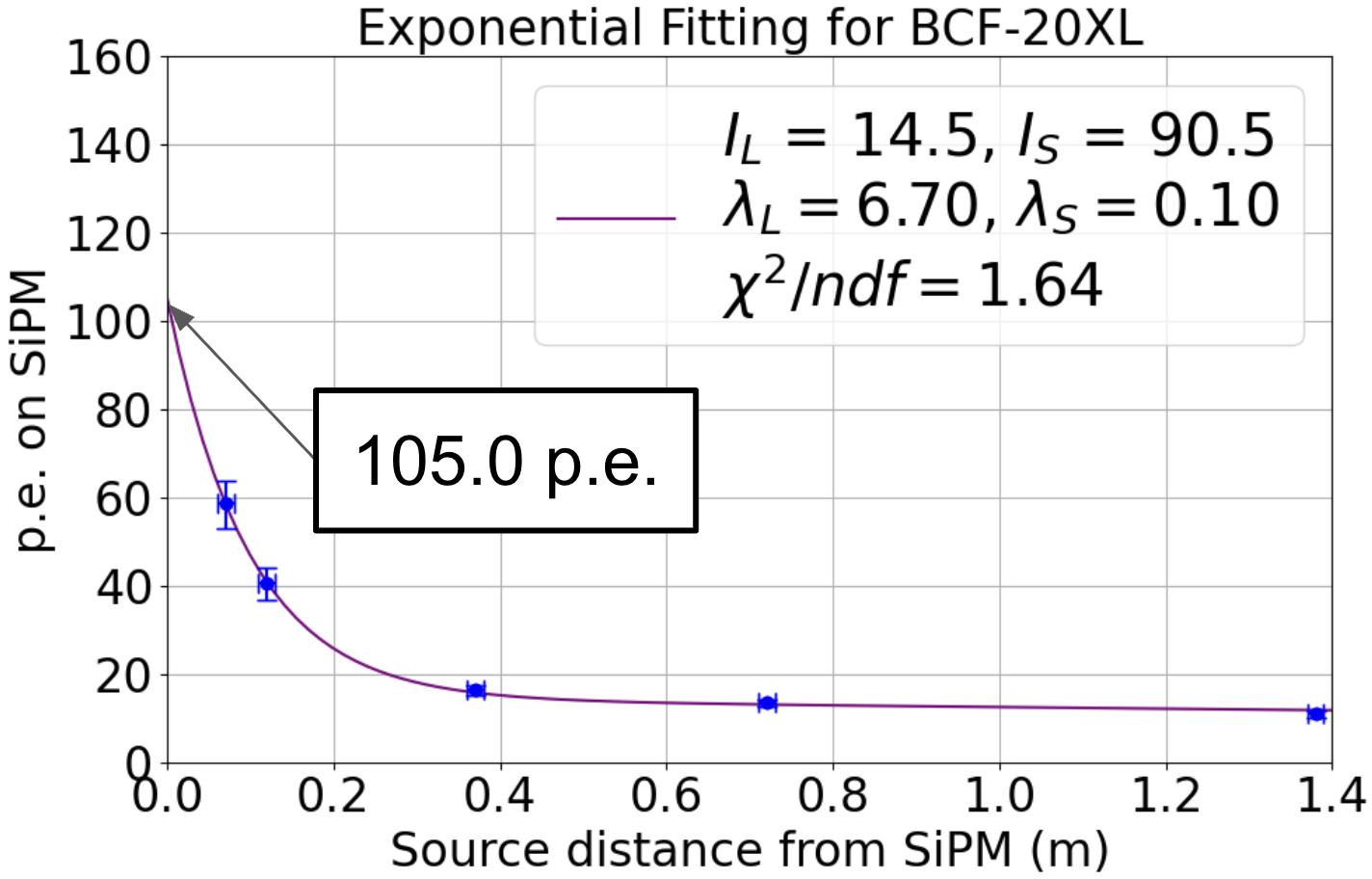}

\caption{Distance dependence of the mean detected p.e. response under \textsuperscript{241}Am $\alpha$ irradiation.
The mean detected p.e. count is shown as  a function of source position, together with its corresponding fit to Eq.~\ref{eq:attenuation_doubleexp_optical}.}
\label{fig: alpha_sum_fitting}
\end{figure}

Figure~\ref{fig: alpha_sum_fitting} summarizes the mean detected p.e. count as a function of fiber length, along with the corresponding fits using the two-component attenuation model in Eq.~\ref{eq:attenuation_doubleexp_optical}.
The resulting $\alpha$ response follows the same qualitative distance dependence observed in the $\beta$ and $\gamma$ measurements.


\subsection{Summary of response to ionizing radiation}
\label{sec:scint_summary}


Table~\ref{tab:bcf20xl_scint_summary} summarizes the BCF-20XL detected photoelectron responses for $\beta$, $\gamma$, and $\alpha$ irradiation.
For each source type, the table gives the extrapolated mean detected p.e. count at $x=0$, the fitted $\lambda_{\mathrm{short}}$, and the fixed $\lambda_{\mathrm{long}}$.
These measurements provide an empirical benchmark for the detected p.e. response and its distance dependence under the specified irradiation and SiPM-readout conditions, relevant to similarly configured detector systems.

As discussed in Section~\ref{sec:optical_setup}, the double-exponential behavior is often attributed to the distinct attenuation of core-guided and cladding-guided light modes, with the cladding-guided component dominating at short distances before attenuating rapidly within the first meter~\cite{Amos1990:fiber-cladding_light, Achenbach:fiber-cladding_light}.
The extracted $\lambda_{\mathrm{short}}$ can therefore be particularly sensitive to the condition of the outer cladding surface and to its optical contact with the surrounding medium in the experimental setup~\cite{Amos1990:fiber-cladding_light, Achenbach:fiber-cladding_light}.
Consequently, numerous studies report $\lambda_{\text{long}}$ by restricting the fitting range to exclude short distances (e.g., 1.0--2.8\,m~\cite{Barbosa2013} and 1.0--4.0\,m~\cite{Zhang:2019LHAASO-AL}).
The spectrophotometer-based measurements in Section~\ref{sec:emission_and_attenuation}, performed with 3.0\,m fiber samples, provide a well-constrained determination of $\lambda_{\text{long}}$.
In the irradiation measurements, however, the fiber length was approximately 1.4\,m, chosen to match the nominal LEGEND-1000 fiber length~\cite{LEGEND:2021bnm}. 
This limited length lacks the necessary leverage to constrain $\lambda_{\text{long}}$ independently.
Therefore, following the approach used in~\cite{Bae:2025ionRad}, we used the $\lambda_{\text{long}}$ value from the 3.0\,m optical-transport measurement in Section~\ref{sec:emission_and_attenuation} and fitted only $\lambda_{\text{short}}$ to describe the measured position dependence.
The uncertainty on $\lambda_{\text{long}}$ is omitted from table~\ref{tab:bcf20xl_scint_summary} to indicate that it was used as a fixed input rather than determined by the fits in this section.


\begin{table}[h!]
    \caption{Summary of the detected scintillation response and distance-dependence fits under $\alpha$, $\beta$, and $\gamma$ irradiation.
    The p.e. counts at $x=0$ are extrapolated values, with $\lambda_{\mathrm{long}}$ fixed at 6.70\,m and $\lambda_{\mathrm{short}}$ fitted.
}
    \centering
    \small
    \begin{tabular}{|>{\raggedright\arraybackslash}m{3.1cm}|>{\centering\arraybackslash}m{3.0cm}|>{\centering\arraybackslash}m{3.0cm}|>{\centering\arraybackslash}m{3.0cm}|}
        \hline
        & Beta from \textsuperscript{90}Sr 
        & Gamma from \textsuperscript{22}Na 
        & Alpha from \textsuperscript{241}Am \\
        \hline
        \makecell[l]{Extrapolated detected \\ p.e. at $x=0$}
        & $(86.8 \pm 5.0)$ p.e.
        & $(74.2 \pm 4.4)$ p.e.
        & $(105.0 \pm 9.4)$ p.e. \\
        \hline
        \makecell[l]{$\lambda_{\text{long}}$}
        & $6.70$ m
        & $6.70$ m
        & $6.70$ m \\
        \hline
        \makecell[l]{$\lambda_{\text{short}}$}
        & $(0.14 \pm 0.01)$ m
        & $(0.13 \pm 0.03)$ m
        & $(0.10 \pm 0.01)$ m \\
        \hline
    \end{tabular}
    \label{tab:bcf20xl_scint_summary}
\end{table}




\section{Conclusions}

In this work, we characterized LED-based optical properties and ionizing-radiation-induced scintillation response for BCF-20XL fiber from Luxium Solutions.
In optical characterization, we measured distance-dependent transmitted emission spectra and described the light attenuation with a two-component attenuation model, obtaining $\lambda_{\mathrm{long}}=(6.70\pm0.08_{\mathrm{fit}})\,\mathrm{m}$ and $\lambda_{\mathrm{short}}=(0.16\pm0.02_{\mathrm{fit}} \pm0.07_{\mathrm{sample}})\,\mathrm{m}$.
We further quantified the wavelength dependence of attenuation by extracting $\Lambda_{\text{long}}(\lambda)$ and $\Lambda_{\text{short}}(\lambda)$ across the emission band.

We also measured the detected photoelectron response of BCF-20XL under standard $\alpha$, $\beta$, and $\gamma$ irradiation using SiPM-based readout. 
For the $\beta$ and $\gamma$ measurements, an approximately 1.4\,m fiber was read out at both ends, while the $\alpha$ measurement used end-irradiated fiber segments of different lengths. 
The distance dependence of the detected response was parameterized with the same two-component attenuation model.
Together, these measurements provide optical-characterization and scintillation-response benchmarks for BCF-20XL.

\acknowledgments{
We thank Luxium Solutions for providing BCF-20XL fiber samples.
We also thank Prof.~S.~Sch\"{o}nert, Dr.~P.~Krause, and the group at the Technical University of Munich for providing BCF-20XL fiber samples and for valuable discussions.
This work was supported in part by the University of Texas at Austin and the U.S.\
National Science Foundation under grant PHY-2312278.
}


\appendix
\clearpage
\section{Side-incident blue-light propagation}
\label{app:side_incident_blue}

\paragraph{Geometry and critical angles.}
Figure~\ref{fig:WLS_vs_SciWLS_guiding}(b) illustrates untrapped transmission under side illumination.
For reference, end-face illumination permits direct guiding within the acceptance cone
(\({\rm NA}=\sqrt{n_{\rm core}^2-n_{\rm clad}^2}\approx0.58\), corresponding to a half-angle of approximately \(36^\circ\) in air)~\cite{SalehTeich2019}.
Here, we consider light entering from air through the side of an ideal straight, circular fiber with smooth, concentric interfaces.
Using \(n_{\rm air}=1.00\), \(n_{\rm clad}=1.49\), and \(n_{\rm core}=1.60\) (table~1), the critical angles measured from the local interface normal are
\[
\alpha_{\rm core}
=\arcsin\!\left(\frac{n_{\rm clad}}{n_{\rm core}}\right)
=68.6^\circ,
\qquad
\alpha_{\rm clad}
=\arcsin\!\left(\frac{n_{\rm air}}{n_{\rm clad}}\right)
=42.2^\circ.
\]
Total internal reflection requires an incidence angle exceeding the corresponding critical angle~\cite{SalehTeich2019, Achenbach:fiber-cladding_light}.

\begin{figure}[htbp]
  \centering
  \includegraphics[width=0.98\linewidth]{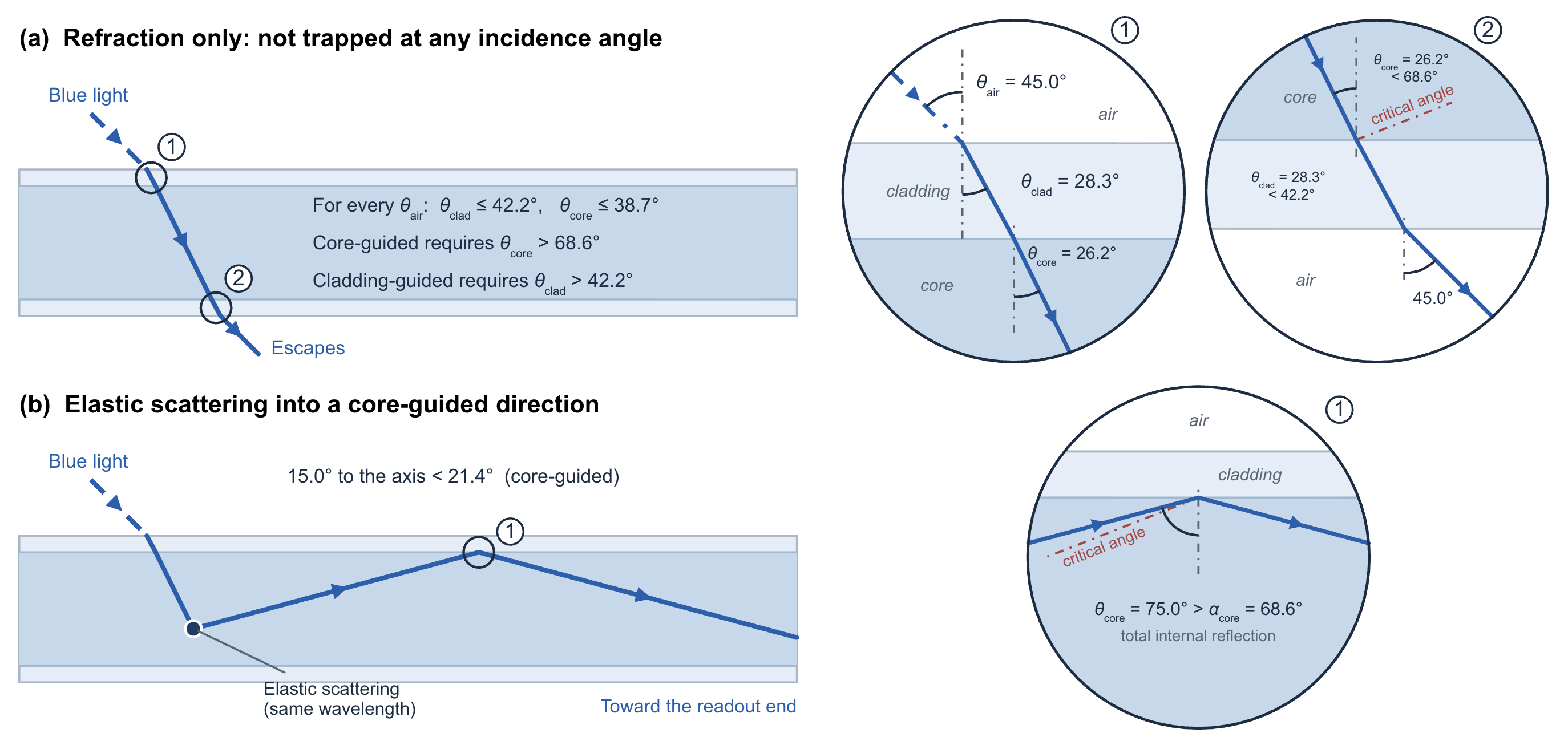}
  \caption{\raggedright
  Side-incident meridional rays at an external angle of \(45^\circ\) to the surface normal.
  Numbered rings correspond to the enlarged views, where the cladding is exaggerated, gray dashed lines indicate normals, and red dashed lines mark critical-angle directions.
  (a) Refraction followed by escape; neither boundary satisfies the trapping condition.
  (b) Elastic scattering into a direction \(15^\circ\) to the axis (\(75^\circ\) to the normal), allowing total internal reflection at the core--cladding boundary.}
  \label{fig:side_incident_blue_paths}
\end{figure}

\paragraph{Meridional rays.}
For a ray in a plane containing the fiber axis, let \(\theta_{\rm air}\), \(\theta_{\rm clad}\), and \(\theta_{\rm core}\) denote the angles to the surface normal in air, cladding, and core.
Snell's law at the two parallel interfaces gives
\begin{equation}
  n_{\rm air}\sin\theta_{\rm air}=n_{\rm clad}\sin\theta_{\rm clad}=n_{\rm core}\sin\theta_{\rm core}.
  \label{eq:side_incidence_angles}
\end{equation}
As \(\theta_{\rm air}\to90^\circ\) (grazing incidence), the internal angles approach their maxima:
\(\theta_{\rm clad}\to42.2^\circ=\alpha_{\rm clad}\) and
\(\theta_{\rm core}\to38.7^\circ<\alpha_{\rm core}\).
Neither exceeds the corresponding critical angle, so refraction alone cannot trap these rays [figure~\ref{fig:side_incident_blue_paths}(a)].

\paragraph{Skew rays.}
Skew rays do not pass through the fiber axis and, unlike meridional rays, are not confined to a single plane~\cite{Achenbach:fiber-cladding_light} [figure~\ref{fig:skew_rays}(a)].
For side-incident light, refraction at the outer surface gives the same angle in the cladding, \(\theta_{\rm clad,1}\), for any plane of incidence;
in the meridional plane, the core--cladding interface is parallel to the outer surface, so the angle at that interface, \(\theta_{\rm clad,2}\), equals \(\theta_{\rm clad,1}\) [figure~\ref{fig:skew_rays}(b)].
However, in the cross-sectional plane, the ray meets the smaller concentric core boundary more obliquely, so that \(\theta_{\rm clad,2}>\theta_{\rm clad,1}\) [figure~\ref{fig:skew_rays}(c)]; skew rays lie between these two cases.
The largest incidence angle at the core--cladding boundary therefore occurs for grazing incidence (\(\theta_{\rm air}\to90^\circ\)) in the cross-sectional plane, for which Snell's law and the concentric geometry give
\begin{equation*}
  n_{\rm clad}\sin\theta_{\rm clad,1}=n_{\rm air},\qquad
  R_{\rm core}\sin\theta_{\rm clad,2}=R_{\rm outer}\sin\theta_{\rm clad,1},\qquad
  n_{\rm core}\sin\theta_{\rm core}=n_{\rm clad}\sin\theta_{\rm clad,2},
\end{equation*}
where \(\theta_{\rm clad,1}\) and \(\theta_{\rm clad,2}\) are the angles in the cladding to the local normals at the outer and core boundaries.
Combining these relations, with \(R_{\rm outer}=0.50\,\mathrm{mm}\) and \(R_{\rm core}=0.47\,\mathrm{mm}\), gives
\begin{equation}
  \theta_{\rm core}\leq\arcsin\!\left(\frac{n_{\rm air}R_{\rm outer}}{n_{\rm core}R_{\rm core}}\right)=41.7^\circ<\alpha_{\rm core}=68.6^\circ .
  \label{eq:skew_max}
\end{equation}
At the cladding--air boundary, as for meridional rays, the incidence angle reaches \(\alpha_{\rm clad}\) only in the grazing limit.
Refraction alone therefore cannot trap side-incident skew rays at either boundary.

{
\begin{figure}[htbp]
  \centering
  \includegraphics[width=0.98\linewidth]{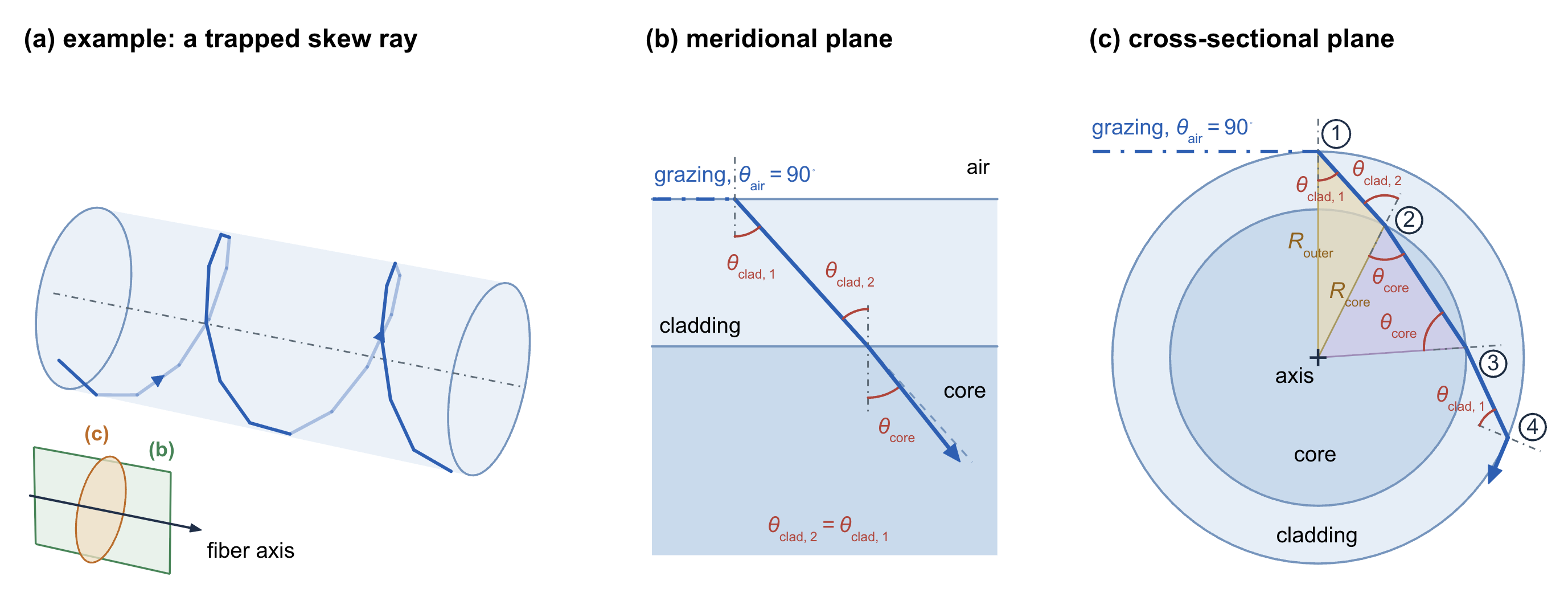}
  \caption{\raggedright
  Skew rays and the limiting side-incidence geometries in a straight, circular fiber with the refractive indices in table~1.
  (a) Example of a trapped skew ray, with the orientations of the planes in (b) and (c).
  (b) Grazing side incidence in the meridional plane: \(\theta_{\rm clad,2}=\theta_{\rm clad,1}\).
  (c) Grazing side incidence in the cross-sectional plane: \(\theta_{\rm clad,2}>\theta_{\rm clad,1}\), the limiting geometry giving the largest angle at the core--cladding boundary.
  The cladding is exaggerated.}
  \label{fig:skew_rays}
\end{figure}
}

\paragraph{Coupling by elastic scattering.}
Scattering, interface defects, or bending can enable coupling beyond the ideal geometry, as illustrated for elastic scattering in figure~\ref{fig:side_incident_blue_paths}(b).
The intrinsic Rayleigh scattering loss of polystyrene, \(55\,\mathrm{dB/km}\) at \(633\,\mathrm{nm}\)~\cite{Kaino1981}, corresponds to a scattering length of approximately \(15.3\,\mathrm{m}\) at \(420\,\mathrm{nm}\) under \(\lambda^{-4}\) scaling.
The scattering probability over a \(1\,\mathrm{mm}\) core path is therefore approximately \(6.5\times10^{-5}\), with only a fraction scattered into guided directions; contributions from interface defects and bending depend on the individual fiber and are not quantified here.

\paragraph{Comparison with WLS absorption.}
In WLS-doped fibers [figure~\ref{fig:WLS_vs_SciWLS_guiding}(a,c)], photons can be trapped following absorption and isotropic re-emission.
The WLS absorption length \(L_{\rm WLS}\) characterizes absorption by the wavelength-shifting fluors and can range from sub-millimetre to millimetre scales within the main absorption band, depending on the material and wavelength~\cite{Bae_PhDthesis, Bae:2026WLSabs}.
Taking \(L_{\rm WLS}=1\,\mathrm{mm}\) as an illustrative value, the Beer--Lambert law~\cite{Swinehart:1962} yields an absorption probability of \(1-e^{-s/L_{\rm WLS}}\approx0.63\) for a core path length \(s=1\,\mathrm{mm}\), roughly four orders of magnitude greater than the estimated Rayleigh-scattering probability at \(420\,\mathrm{nm}\).
Side-incident light entering the doped core is therefore expected to be collected mainly through the WLS process.




\clearpage
\providecommand{\href}[2]{#2}\begingroup\raggedright\endgroup

\end{document}